# Conduction Electron Spin-Flipping at Sputtered Co$_{90}$Fe$_{10}$/Cu Interfaces.

H.Y.T.Nguyen, R. Acharyya, W.P. Pratt Jr., and J. Bass
Department of Physics and Astronomy, Michigan State University, East Lansing, MI 48824

Abstract
From measurements of the current-perpendicular-to-plane (CPP) magnetoresistance of ferromagnetically coupled [Co$_{90}$Fe$_{10}$/Cu]$_n$ multilayers, within sputtered Permalloy-based double exchange biased spin-valves, we determine the parameter $\delta_{Co(90)Fe(10)/Cu} = 0.19 \pm 0.04$ that sets the probability $P$ of spin-flipping at a Co$_{90}$Fe$_{10}$/Cu interface via the equation $P = 1 - \exp(-\delta)$.

I. Introduction.

One of the few fundamental parameters in electrical resistance of ferromagnetic/non-magnetic (F/N) metallic multilayers measured with current-flow perpendicular to the layer planes (CPP) [1-3] that has not yet been determined for several different F/N pairs is the probability $P_{F/N}$ of spin-flipping at the F/N interfaces. This probability is related to a parameter $\delta_{F/N}$ via the equation $P_{F/N} = 1 - \exp(-\delta_{F/N})$. We have argued elsewhere [1] that most of the few published claims of measurements of $\delta_{F/N}$ are unreliable. This unreliability led us to recently propose a new technique for measuring $\delta_{F/N}$ [2], and to test the technique by applying it to F/N = Co/Cu. The advantage of Co was that its bulk spin-flipping parameter, the spin-diffusion length $l_{sf}^{Co}$, is believed to be unusually long, $l_{sf}^{Co} \sim 60$ nm [4,5] at our measuring temperature T = 4.2K, letting us focus on spin-flipping due to $\delta_{F/N}$. Happily, our new value of $\delta_{Co/Cu} = 0.33^{+0.3}_{-0.8}$ agreed to within mutual uncertainties with those estimated from other less direct techniques (for details see [2]).

In the present paper, we apply the same technique to a second F/N pair: Co$_{90}$Fe$_{10}$/Cu. We chose Co$_{90}$Fe$_{10}$ because its lattice is close to that of Cu, and its CPP-MR parameters are generally similar to those of Co, except that its $l_{sf}^{Co(90)Fe(10)} \sim 12$ nm [6] is much shorter than that of Co, but still long enough to not mask additional spin-flipping due to $\delta_{Co(90)Fe(10)/Cu}$. The question of interest is whether $\delta_{Co(90)Fe(10)/Cu}$ differs significantly from $\delta_{Co/Cu}$. Hereafter we simplify Co$_{90}$Fe$_{10}$ to just CoFe.

The technique involves analyzing data on three different sets of samples using the theory of Valet and Fert (VF). The first set is simple [CoFe/Cu]$_n$ multilayers, where $n$ is the number of repeats. VF theory shows that these previously measured data [7] should be insensitive to $\delta$. The second set is symmetric [FeMn/CoFe(t)/Cu/CoFe(t)] exchange-biased spin-valves (EBSVs), where FeMn is an antiferromagnet used to pin the magnetization of the adjacent CoFe layer (pinned layer) so that it reverses at a much higher magnetic field H than does the other CoFe layer (free layer)[6]. VF theory shows that its data are moderately sensitive to $\delta$. For internal consistency, we replace previously published data for Co$_{91}$Fe$_9$/Cu EBSVs [6] with new data for Co$_{90}$Fe$_{10}$/Cu. By itself, this replacement makes only small changes in parameters. The third set is symmetric double EBSVs (DEBSVs) of the form [FeMn/Py/Cu/[Co$_{90}$Fe$_{10}$/Cu]$_n$/Co$_{90}$Fe$_{10}$/Cu/Py/IrMn], where the thickness of the Cu layers between the Co$_{90}$Fe$_{10}$ ones is fixed to give ferromagnetic coupling between the adjacent CoFe layers, the thickness of the Co$_{90}$Fe$_{10}$ is chosen to be several times shorter than the bulk $l_{sf}^{CoFe}$, and Py = Ni$_{1-x}$Fe$_x$ with x ~ 0.2. VF theory shows that data for such a DEBSV are sensitive to $\delta$.

The paper is organized as follows. In section II, we explain in detail the technique for measuring $\delta_{F/N}$. In section III we briefly describe our experimental techniques and samples. In section IV we present the parameters that we predetermined for our samples, our new data, and our analysis. Section V contains a summary and conclusions.

II. The technique.

In the Valet-Fert (VF) theory of CPP-MR [8], an F/N pair of metals has eight parameters: the resistivity $\rho_N$, and the spin-diffusion length, $l_{sf}^N$, within the N-metal; the enhanced resistivity, $\rho_F^*$, the bulk scattering asymmetry, $\beta_F$, and the spin-diffusion length, $l_{sf}^F$, within the F-metal; and the enhanced interface specific resistance, $AR_{F/N}^*$, the interfacial scattering asymmetry, $\gamma_{F/N}$, and the interfacial spin-flipping parameter, $\delta_{F/N}$. Two of these, $\rho_N$ and $\rho_F = [1-(\beta_F)^2]\rho_F^*$, can be measured using the Van der Pauw (VDP) technique on separately prepared thin N- and F-films, leaving six. For our Cu, $l_{sf}^{Cu} \geq 500$ nm [1] is so much longer than our Cu layer thickness that it can be approximated as $\infty$. The four remaining parameters, other than $\delta_{F/N}$, are first approximated from CPP-MR measurements on the first two kinds of samples noted above, assuming that $\delta_{CoFe/Cu} = 0$. For the EBSVs, we use new data from the same Co$_{90}$Fe$_{10}$ target as used in [7] and in the rest of the present paper. With these approximate parameters in hand, we then turn to the DEBSVs of the form FeMn/Py/Cu/X/Cu/Py/FeMn with X = [CoFe/Cu]$_n$CoFe. To make X reverse as a single entity, we choose the thickness of the Cu-layers = 1.4 nm to give ferromagnetic coupling between the neighboring F-layers. We then measure, as a function of $n$, the change in specific resistance, $A\Delta R = AR(AP) - AR(P)$, between the states where the magnetization of X is oriented anti-parallel (AP) or parallel (P) to the common direction of the moments of the pinned Py (See, e.g., Fig. 1 in [2]). By increasing the number of interfaces, we make $A\Delta R$ sensitive to the value of $\delta_{CoFe/Cu}$. The DEBSV also gives an $A\Delta R$ about twice that for a single EBSV, and the symmetry of the DEBSV about its middle lets us do the numerical calculation for only half of the multilayer, thereby greatly simplifying the computer program and analysis. Starting from the parameters determined from the first two sets of samples assuming $\delta_{CoFe/Cu} = 0$, we fit all three sets of data self-consistently to determine $\delta_{CoFe/Cu} \neq 0$.

III. Preparing and Measuring Samples.

Our sputtering and measuring techniques have been described elsewhere [9]. All of the multilayers were sputtered in an ultra-high-vacuum compatible, six-target, sputtering system with background vacuum ≤ 2 x 10$^{-8}$ Torr and purified Ar as the sputtering gas. To achieve uniform CPP current flow through the sample, the multilayers were sandwiched between ~ 1.1 nm wide, 150 nm thick, crossed Nb strips, which become superconducting at our measuring temperature of 4.2K. The DEBSV structure was (layer thicknesses in nm). Nb(150)/Cu(10)/FeMn(8)/Py(6)/Cu(10)/[CoFe(3)/Cu(1.4)]$_n$CoFe(3)/Cu(10)/Py(6)/FeMn(8)/Cu(10)/Nb(150). As shown in refs. [2,6], 8 nm of FeMn is thick enough to give strong pinning. 6 nm of Py is comparable to $l_{sf}^{Py}$ = 5.5 ± 1 nm [1,10]. The Cu(10) layer above the bottom Nb strip helps the FeMn to grow in the proper structure for good pinning. The Cu(10) layer below the top Nb makes the sample centrosymmetric. This centrosymmetry lets us evaluate numerically the VF expression for $A\Delta R$ for only half of the sample, thereby simplifying the calculation. To keep $8t_{CoFe} \leq 2 l_{sf}^F$, we chose $t_{CoFe} = 3$ nm.

IV Parameters, Data, and Analysis.

To check the Cu thickness $t_{Cu}$ needed for ferromagnetic coupling, we used the data in [11] as a guide and measured the in-plane saturation field, $H_{sat}$, for 1 nm ≤ $t_{Cu}$ ≤ 1.5 nm. The minimum $H_{sat}$ (maximum ferromagnetic coupling) occured at $t_{Cu} = 1.4$ nm, which we chose.

For the VF calculation, we first adopt our previously published values of the bulk and interface properties of metals other than CoFe and Cu [2,6,12]: $\rho_{FeMn} = 875 \pm 50$ nΩm; $AR_{Nb/FeMn} = 1.0 \pm 0.6$ fΩm$^2$; $AR_{Nb/CoFe} = 3.5 \pm 0.5$ fΩm$^2$; $AR_{FeMn/CoFe} = 0.95 \pm 0.1$ fΩm$^2$; $AR_{FeMn/Py} = 1.0 \pm 0.4$ fΩm$^2$; $\rho_{Py} = 123 \pm 40$ nΩm; $\beta_{Py} = 0.76 \pm 0.07$; $l_{sf}^{Py}$ = 5.5 ± 1 nm; $2AR^*_{Py/Cu} = 1.0 \pm 0.08$ fΩm$^2$; $\gamma_{Py/Cu} = 0.7 \pm 0.1$. For $\rho_{Cu}$, we checked with new VdP measurements that our published value, $\rho_{Cu} = 4.5$ nΩm, is still valid. For $\rho_{CoFe}$, a combination of new and old VdP measurements on 200 nm thick films led us to choose a new value of $\rho_{CoFe} = 75 \pm 5$ nΩm. This value is consistent with, but has lower uncertainty than, the earlier assumed 70 ± 10 nΩm from measurements on Co$_{91}$Fe$_9$ films [6]. We also kept $l_{sf}^{CoFe} = 12 \pm 1$ nm [6]. To find the rest of our starting parameters for CoFe and CoFe/Cu interfaces we used VF theory, assuming $\delta_{CoFe/Cu} = 0$, to fit a combination of the slopes of $A\Delta R$ vs $n$ (1.80 ± 0.05 fΩm$^2$) and $\sqrt{AR(AP)A\Delta R}$ vs $n$ (1.34 ± 0.04 fΩm$^2$) for our published data on simple [CoFe/Cu]$_n$ multilayers [7], with new data on [FeMn/CoFe/Cu/CoFe] EBSVs (Fig. 1). This joint fit gave the parameters listed in column 3 of Table 1, which give the solid curve in Fig. 1 and the dashed curve in Fig. 2. As required for internal consistency, the dashed curve in Fig. 2 fits the data at $n = 0$. But it falls above the data for $n > 2$, suggesting the need for $\delta \neq 0$.

We end with a complete fit to all three data sets with $\delta \neq 0$. Simply adding to the parameters of column 3 a value of $\delta_{CoFe/Cu} = 0.19$ gives a good fit to the DEBSV data. However, adding this non-zero $\delta_{CoFe/Cu}$ worsens the fit to the EBSV data of Fig. 1. To fit those data requires an increase in $\beta_{CoFe}$. To then fit the slopes of the simple multilayer data, this increase in $\beta_{CoFe}$ causes small decreases in $\gamma_{CoFe/Cu}$ and $2AR^*_{CoFe/Cu}$. These decreases reduce the intercept in Fig. 2, bringing the best fit curve below the data for small $n$. To improve the fits for small $n$ in Fig. 2, we increased the slopes of the simple multilayer data to 1.85 fΩm$^2$ for $AR(AP)$ vs $n$ and to 1.38 fΩm$^2$ for $\sqrt{AR(AP)A\Delta R}$ vs $n$, each 1 standard deviation above the best fits given above. Then, using these slopes, plus the value of $A\Delta R$ in Fig. 2 for $n = 7$ (chosen as the average over $n = 6$-8), and the EBSV data in Fig. 1, we adjust $\beta_{CoFe}$ and $\delta_{CoFe/Cu}$ to give the best joint fit. This process gives the parameters in column 4 of Table 1 and the solid curves in Figs. 1 and 2 (in Fig. 1, the curves for the parameters of columns 3 and 4 are indistinguishable). These parameters are our best estimates. Except for $\delta$, they are consistent with our previously published values in column 2 [7], to within mutual uncertainties.

V. Summary and Conclusions.

We measured the current-perpendicular-to-plane (CPP) magnetoresistance of three sets of sputtered multilayers: (a) simple Co$_{90}$Fe$_{10}$/Cu multilayers; (b) symmetric [FeMn/$Co_{90}Fe_{10}$(t)/Cu/$Co_{90}Fe_{10}$(t)] exchange-biased spin-valves (EBSVs); and (c) ferromagnetically coupled [$Co_{90}Fe_{10}$/Cu]$_n$ multilayers, within sputtered Permalloy-based double exchange biased spin-valves (DEBSVs). From these measurements we derived the bulk scattering parameters for $Co_{90}Fe_{10}$ and the interface scattering parameters for $Co_{90}Fe_{10}$/Cu interfaces, as given in column 4 of Table I. Our main new result is the non-zero value of $\delta_{Co(90)Fe(10)/Cu} = 0.19 \pm 0.04$ that sets the probability $P$ of spin-flipping at a $Co_{90}Fe_{10}$/Cu interface via the equation $P = 1 - \exp(-\delta)$. This parameter is comparable to, but smaller than, the parameter for Co/Cu interfaces: $\delta_{Co/Cu}$ $0.33^{+0.3}_{-0.8}$ [2]. The other parameters in column 4 are consistent, to within mutual uncertainties, with our previously published values given in column 2.

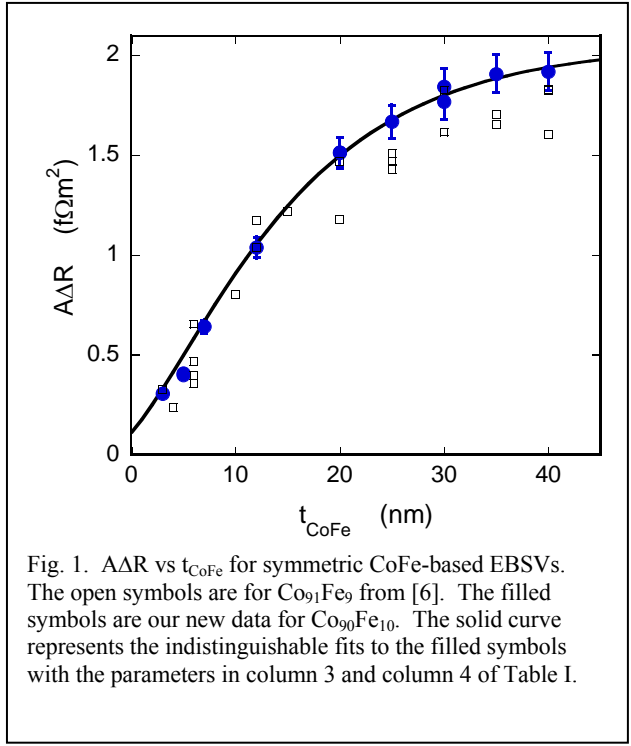

Fig. 1. $A\Delta R$ vs $t_{CoFe}$ for symmetric CoFe-based EBSVs. The open symbols are for $Co_{91}Fe_9$ from [6]. The filled symbols are our new data for $Co_{90}Fe_{10}$. The solid curve represents the indistinguishable fits to the filled symbols with the parameters in column 3 and column 4 of Table I.

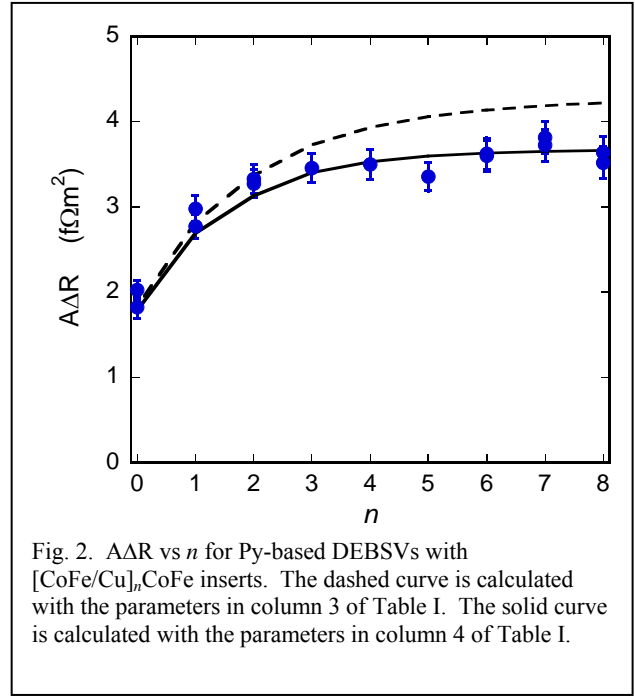

Fig. 2. $A\Delta R$ vs $n$ for Py-based DEBSVs with $[CoFe/Cu]_nCoFe$ inserts. The dashed curve is calculated with the parameters in column 3 of Table I. The solid curve is calculated with the parameters in column 4 of Table I.

Acknowledgments: This work was supported in part by NSF grant DMR 08-04126 and a grant from the Korea Institute for Science and Technology (KIST).

Table I. Parameters for three different circumstances. The column 'Kim-Lee' lists the parameters from ref. [7]. The column '$Co_{90}Fe_{10}$:$\delta = 0$' lists the parameters determined from Fig. 1 in ref. [7] and Fig. 1 in the present paper, assuming $\delta = 0$. The column '$Co_{90}Fe_{10}$: $\delta \neq 0$' lists our self-consistent 'best fit' parameters including the data in Fig. 2.

|  | Kim-Lee | $Co_{90}Fe_{10}$:$\delta=0$ | $Co_{90}Fe_{10}$:$\delta\neq0$ |
|---|---|---|---|
| $\rho_F$(n$\Omega$m) | 70±10 | 75±5 | 75±5 |
| $\beta_F$ | 0.65±0.05 | 0.67±0.04 | 0.71±0.04 |
| $l_{sf}$(nm) | 12±1 | 12±1 | 12±1 |
| 2AR*$_{F/Cu}$(f$\Omega m^2$) | 1.1±0.2 | 1.0±0.1 | 0.97±0.1 |
| $\gamma_{F/Cu}$ | 0.8±0.1 | 0.78±0.05 | 0.76±0.05 |
| $\delta$ | 0 | 0 | 0.19±0.04 |